\documentclass{emulateapj}
\usepackage{psfig}

\newcommand{\etal}{\mbox{et al.}}
\newcommand{\ergcms}{erg cm$^{-2}$ s$^{-1}$}
\newcommand{\ergs}{erg s$^{-1}$}
\newcommand{\phcms}{ph cm$^{-2}$ s$^{-1}$}
\newcommand{\degree}{$^\circ$}
\newcommand{\msun}{$M_{\odot}$}

\newcommand{\chandra}{{\it Chandra}}

\newcommand{\xmm}{{\it XMM-Newton}}

\newcommand{\wdpulse}{\mbox{CXO J164710.2--455216}}

\shortauthors{Muno \etal}
\shorttitle{A Neutron Star in Westerlund 1}

\begin{document}

\title{A Neutron Star with a Massive Progenitor in Westerlund 1}

\author{
Michael P. Muno,\altaffilmark{1,2}
        J. Simon Clark,\altaffilmark{3}
	Paul A. Crowther,\altaffilmark{4}
	Sean M. Dougherty,\altaffilmark{5}
	Richard de Grijs,\altaffilmark{4}
	Casey Law,\altaffilmark{6}
	Stephen L. W. McMillan,\altaffilmark{7}
	Mark R. Morris,\altaffilmark{2}
	Ignacio Negueruela,\altaffilmark{8}
	David Pooley,\altaffilmark{9,10}
	Simon Portegies Zwart,\altaffilmark{11,12}
	Farhad Yusef-Zadeh\altaffilmark{6}
}

\altaffiltext{1}{Department of Physics and Astronomy, University of California,
Los Angeles, CA 90095; mmuno@astro.ucla.edu}
\altaffiltext{2}{Hubble Fellow}
\altaffiltext{3}{Department of Physics \& Astronomy, The Open University, 
Walton Hall, Milton Keynes, MK7 6AA, UK}
\altaffiltext{4}{Department of Physics \& Astronomy, The University of 
  Sheffield, Hicks Building, Hounsfield Road, Sheffield S3 7RH, U.K.}
\altaffiltext{5}{National Research Council, Herzberg Institute of Astrophysics,
   Dominion Radio Astrophysical Observatory, PO Box 248, Penticton BC. V2A 6K3}
\altaffiltext{6}{Department of Physics and Astronomy, Northwestern University, 
   Evanston, IL 60208, USA}
\altaffiltext{7}{Department of Physics, Drexel University, Philadelphia, PA 
 19104.}
\altaffiltext{8}{Ram\'on y Cajal Fellow, Dpto. de F\'{i}sica,
       Ingenier\'{i}a de Sistemas y Teor\'{\i}a de la Se\~{n}al, Universidad de
       Alicante, Apdo. 99 E03080, Alicante, Spain}
\altaffiltext{9}{Chandra Fellow}
\altaffiltext{10}{Astronomy Department, University of California at Berkeley, 
601 Campbell Hall, Berkeley, CA 94720, USA}
\altaffiltext{11}{Astronomical Institute 'Anton Pannekoek'
        Kruislaan 403, 1098SJ Amsterdam, the Netherlands}
\altaffiltext{12}{Section Computational Science 
        Kruislaan 403, 1098SJ Amsterdam, the Netherlands}

\begin{abstract}
We report the discovery of an X-ray pulsar in the young, massive
Galactic star cluster Westerlund 1. We detected a coherent signal from the 
brightest X-ray source in the cluster, \wdpulse, during two \chandra\ 
observations on 2005 May 22 and June 18. The period of the pulsar is 
10.6107(1) s. We place an upper limit to the period derivative of 
$\dot{P}<2\times10^{-10}$ s s$^{-1}$, which implies that the spin-down
luminosity is $\dot{E} \le 3\times10^{33}$ \ergs. The X-ray luminosity 
of the pulsar is 
$L_{\rm X} \approx 3^{+10}_{-2}\times10^{33} (D/5~{\rm kpc})^2$~\ergs,
and the spectrum can be described by a 
$kT = 0.61^{+0.02}_{-0.02}$ keV blackbody with a radius of 
$R_{\rm bb} = 0.27\pm0.03 (D/{5~{\rm kpc}})$ km. Deep infrared observations
reveal no counterpart with $K$$<$18.5, ruling out a binary companion
with $M$$>$1\msun. Taken together, the properties of the pulsar
indicate that it is a magnetar. The rarity of slow X-ray pulsars and the 
position of \wdpulse\ only 1.6\arcmin\ from the core of Westerlund 1
indicates that it is a member of the cluster with $>$99.97\% confidence.
Westerlund 1 contains 07V stars with initial masses $M_i$$\approx$35\msun\
and $>$50 post-main-sequence stars that indicate the 
cluster is 4$\pm$1 Myr old. Therefore, the progenitor to this pulsar had 
an initial mass $M_i$$>$40\msun. This is 
the most secure result among a handful of observational limits to the 
masses of the progenitors to neutron stars.
\end{abstract}


\section{Introduction}

Most of our knowledge about the masses of the progenitors to neutron stars 
is based on theoretical calculations \citep[e.g.,][]{heg03}. Quantitative 
observational constraints are difficult to establish. The few previous,
tentative estimates have relied on inferring these masses from traces of 
the interactions of the progenitors with their 
surroundings \citep{nom82,mac89,gae05}, on developing scenarios
by which individual accreting X-ray binaries could have 
formed \citep[e.g.,][]{eh98,wl99}, or on 
demonstrating that the progenitor was a member of a population of coeval 
stars with well-determined masses \citep[e.g.,][]{fuc99,vrb00,fig05,pel05}.
The first two methods remain uncertain because of their dependence on
assumptions in the relevant models. The third class of results are
potentially the most reliable, although there is sometimes debate
over the whether the neutron stars and 
the stellar associations are related \citep[e.g.,][]{cam05,mcg05}. 

Here we report the discovery of a neutron star in 
\chandra\ observations of the Galactic star cluster Westerlund 1. 
%($\alpha$,$\delta$ = 16 47 03.7, --45 51 00; J2000). 
The cluster contains an exceptional population of $>$50 massive, 
post-main-sequence stars that are only 4$\pm$1 Myr old \citep{wes87,cla05}. 
Its total mass of $>$$10^5$\msun\ is contained in a region only $\approx$9 pc 
across, which suggests that the stars were born 
in a single episode of star formation \citep{cla05}. This makes it an 
ideal cluster for placing limits on the initial masses of progenitors to 
compact objects.

%%%%%%%%%%%%%%%%%%%%%%%%%%%%%%%%%%%%%%%%%%%%%%%%%%%%%%%%%%%%%%%
%% the X-ray image
\begin{figure}
\centerline{\psfig{file=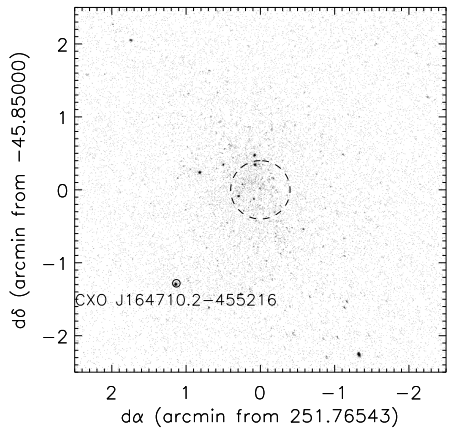,width=0.8\linewidth}}
\caption{The central 5\arcmin$\times$5\arcmin\ of the
image we obtained of Westerlund 1 with \chandra. The image is 
a composite of two exposures, taken on 2005 May 22 and June 18.
The count image
has been corrected to account for the varying exposure across
the image.
We have indicated the 
centroid and core-radius of the cluster derived from the locations of 
the X-ray sources using the dashed line. The location of the pulsar
is indicated with the solid circle.
}
\label{fig:img}
\end{figure}

%%%%%%%%%%%%%%%%%%%%%%%%%%%%%%%%%%%%%%%%%%%%%%%%%%%%%%%%%%%%%%%
%% the radial distribution
\begin{figure}
\centerline{\psfig{file=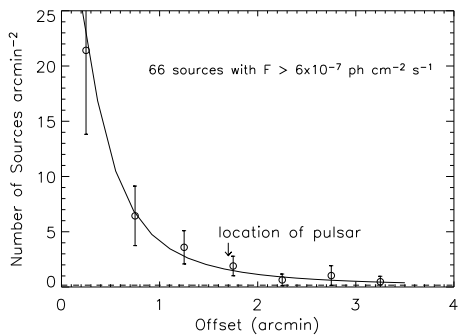,width=0.8\linewidth}}
\caption{
The binned radial distribution of point sources brighter than 
our 50\% completeness limit. The distribution is computing assuming that
its centroid is at $\alpha$,$\delta$ = 16 47 03.7, --45 51 00 (J2000).
The solid line is the best-fit Lorentzian model of the {\it unbinned}
distribution, which has a width of $\theta_0 = 0.4\pm0.1$ arcmin and
a central surface density of $\rho_0 = 33^{+16}_{-10}$ sources 
arcmin$^{-2}$.}
\label{fig:dist}
\end{figure}

\section{Observations and Data Analysis}

We observed Westerlund 1 with the {\it Chandra X-ray Observatory} Advanced 
CCD Spectrometer Spectroscopic array (ACIS-S)\cite{wei02} on two 
occasions: 2005 May 22 for 18 ks 
(sequence 5411) and 2005 June 18 for 
42 ks (sequence 6283). 
The event lists were reduced using the standard tools and
techniques described on the web site of the 
{\it Chandra X-ray Center} (CXC).

We used the wavelet-based algorithm 
{\tt wavdetect} \citep{free02} to identify 238 individual point-like 
X-ray sources in the entire image (which covered approximately four times 
the area displayed in Fig.~\ref{fig:img}). Half of these sources 
are located in the central portion of the image shown.
Above our completeness limit of 
$6\times10^{-7}$~\phcms\ (0.5--8.0 keV), we find that the central 
surface density of X-ray sources is $33^{+16}_{-10}$ arcmin$^{-2}$.
For comparison, observations of the Galactic plane at 
$l = 28^\circ$ and $b = 0.2^{\circ}$ reveal only 0.17$\pm$0.04 
sources arcmin$^{-2}$ at this flux level \citep[e.g.,][]{ebi01}. The surface 
density of sources can be modeled as a Lorentzian function with half-width 
$0.4$$\pm$$0.1$ arcmin (Fig.~\ref{fig:dist}). Therefore, even without 
considering the properties of individual
X-ray sources, any one located within 1\farcm7 of the core of 
Westerlund 1 has a $>$90\% chance of being a cluster member.
%and those within 3\arcmin\ have a 50\% chance of being cluster members.
%The distribution of X-ray sources is similar to that of cluster stars
%observed in optical and infrared images \citep{cla05}.

Most of the X-ray sources have Wolf-Rayet or O (WR/O) stars as counterparts in 
optical or infrared images, and are probably colliding-wind binaries.
However, some of the WR/O stars could be in binaries with accreting 
compact objects, and X-ray sources without obvious stellar counterparts could 
be isolated pulsars. We did not find any obvious candidates for 
accreting black holes (J. S. Clark \etal, in prep).

%%%%%%%%%%%%%%%%%%%%%%%%%%%%%%%%%%%%%%%%%%%%%%%%%%%%%%%%%%%%%%%
%% periodograms. 
\begin{figure}
\centerline{\psfig{file=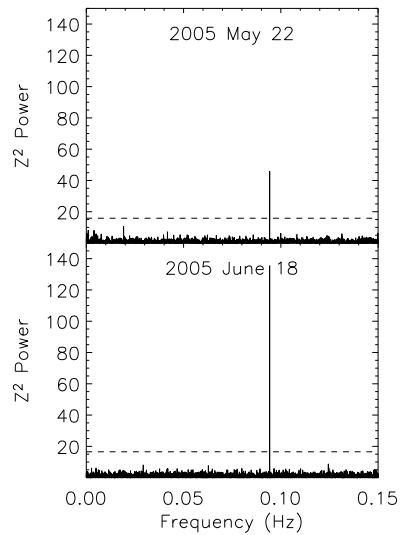,width=0.6\linewidth}}
\caption{
The periodogram of \wdpulse, which is the only source to exhibit 
a significant periodic signal. The dashed lines denote the powers
above which there is a $<$1\% chance that noise would produce signals
that large. The signal from \wdpulse\ produced $Z^2_1 = 58.3$ from 398 
counts in the May observation, and $Z^2_1 = 135$ from 857 events in the June 
observation. The periodicity was significant at the $>$8$\sigma$ 
level. The profile of the signal was sinusoidal, and the fractional
rms amplitudes were 53.0(1)\% and 54.91(3)\%.
}
\label{fig:periodogram}
\end{figure}

The best means of identifying neutron stars is to search for 
rotationally-modulated X-ray emission.  Therefore, for the brightest
sources, we adjusted the arrival times of their photons 
to the Solar System barycenter and computed Fourier periodograms using the 
Rayleigh statistic \citep[$Z_1^2$;][]{buc83}. 
The individual X-ray events were recorded with a time resolution of 3.2~s, so 
the Nyquist frequency was $\approx$0.15~Hz, which represents the limit 
below which our sensitivity could be well-characterised. 
However, we computed the periodogram using a maximum frequency of 
$\approx$0.6 Hz, to take advantage of the limited sensitivity to higher 
frequency signals, and to ensure that any observed signal was not an alias. 
We searched the two observations separately, so that 
a search for signals from a single source required 3221 independent 
trials for the May observation, and 6239 trials for the June 
observation. The corresponding single-source detection threshold powers
for 99\% confidence in each observation are $Z_1^2$=12.7 and 13.3, 
respectively, where the powers have been normalised so that Poisson 
noise produces power with a mean value of 1. This power can be related to 
the root-mean-squared 
(rms) amplitude of a sinusoidal signal by $A = (2Z_1^2/N_\gamma)^{1/2}$,
where $N_\gamma$ is the number of photons from the source. 
A fully-modulated signal would have an amplitude of 0.71, so the minimum 
number of counts required to detect a signal are $N_\gamma$=51 and 53, 
respectively. We searched 8 sources
above this count limit in the May observation, and 16 sources in the 
June observation.

%%%%%%%%%%%%%%%%%%%%%%%%%%%%%%%%%%%%%%%%%%%%%%%%%%%%%%%%%%%%%%%
%% Spectrum
\begin{figure}
\centerline{\psfig{file=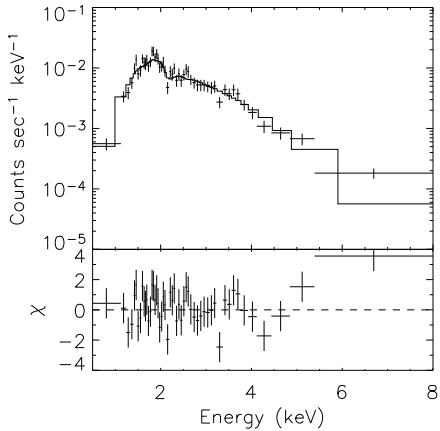,width=0.7\linewidth}}
\caption{
The X-ray spectrum of \wdpulse. The top panel displays the spectrum in 
detector counts as a function of energy, so the shape of the source 
spectrum is convolved with the detector response. Several models 
reproduced the data equally well (see text).
 The bottom panel contains the difference between the
data and the best-fit black body model, divided by the 
Poisson uncertainty on the data points.}
\label{fig:spectrum}
\end{figure}

One periodic signal 
from the brightest X-ray source in the field, \wdpulse\ 
($\alpha,\delta$ = 251.79250,--45.87136 [J2000], $\pm$0\farcs3 with 
90\% conf.), significantly exceeded the 
expected noise in both observations (Fig.~\ref{fig:periodogram}).
We refined an initial estimate of the period by computing pulse profiles 
from non-overlapping 5000 s intervals during each observation, and modelling 
the differences between the assumed and measured phases using first-and 
second-order polynomials. The reference epochs of the pulse maxima for the two 
observations were 53512.860265(4) and 53539.67325(2) (MJD, TBD).
The best-fit periods were 10.6112(4) s and 
10.6107(1) s for 2005 May and June, respectively. This placed a limit 
on the difference in period between the two observations of 
$\Delta P < 0.5$ ms, or on the period derivative of 
$\dot{P}<2\times10^{-10}$ s s$^{-1}$. From the quadratic fits 
to the phases during the individual May and June 
observations, we found $\Delta P < 2$ and $< 0.4$ ms
($\dot{P}<9\times10^{-8}$ s s$^{-1}$ and $<1\times10^{-8}$ s s$^{-1}$), 
respectively. Unfortunately, there was a $\approx$2 cycle ambiguity when
trying to predict the phases over the month interval between observations,
so the period cannot be refined further using the current observations. 
Nonetheless, the stability of the signal from \wdpulse\ 
suggests that it is produced by the rotation of neutron star.

We extracted spectra of \wdpulse\ using standard tools and the 
acis\_extract routine.\footnote{http://www.astro.psu.edu/xray/docs/TARA/}
We found no evidence that the intensity of the source changed 
between May and June, so we combined the spectra obtained from both 
observations. We modeled the spectrum using
XSPEC version 12 \citep{arn96}. All uncertainties were computed using 
1$\sigma$ confidence intervals ($\Delta\chi^2$$=$1). The spectrum could 
be described equally well
($\la$10\% chance probability) by black body, bremsstrahlung, or power 
law continuum models that were absorbed and scattered by the 
interstellar medium (ISM). 
To facilitate comparisons with other isolated neutron stars, 
we report the parameters of black body model ($\chi^2/\nu = 63.8/49$),
for which we derive an absorption column of
$N_{\rm H} = 1.2\pm0.1 \times 10^{22}$ cm$^{-2}$, 
a best-fit temperature of $kT = 0.61\pm0.02$ keV, and an apparent radius of 
$R_{\rm bb} = 0.27\pm0.03 (D/{5~{\rm kpc}})$ km. 
%
%An optically-thin bremsstrahlung model ($\chi^2/\nu = 57.8/49$) implied
%$N_{\rm H} = 1.9\pm0.1 \times 10^{22}$ cm$^{-2}$, 
%had a best-fit temperature of 
%$kT = 1.5\pm0.1$ keV, and had an emission measure
%of $\int n_{\rm e} n_{\rm I} dV = 5\pm1\times10^{46} 
%(D/{5~{\rm kpc}})^2$ cm$^{-3}$. 
%A power-law continuum model ($\chi^2/\nu = 60.1/49$) implied
%$N_{\rm H} = 2.6\pm0.2 \times 10^{22}$ cm$^{-2}$, 
%had a best-fit photon index $\Gamma = 3.8\pm0.2$,  and had
%a normalization of $1.6\pm0.5 \times 10^{-3}$ photons keV$^{-1}$
%cm$^{-2}$ s$^{-1}$ at 1 keV.

%% If we apply a bremmstrahlung or power law model, we find that 
Comparing the three continuum models, we found that the interstellar 
extinction toward the source is
$N_{\rm H} = (1.2-2.6)\times10^{22}$ cm$^{-2}$. This 
is consistent with the range measured for other cluster members, 
$N_{\rm H} = [1.4-2.9]\times10^{22}$ cm$^{-2}$, where the dispersion 
can be accounted for by variable extinction caused by a foreground 
molecular cloud (see Clark \etal\ 2005).
The observed flux was $F_{\rm X} = 2.4^{+0.2}_{-0.6} \times 10^{-13}$ \ergcms\ 
(0.5--8.0 keV). We de-reddened the three models, and found
$L_{\rm X} = 3^{+10}_{-2}\times10^{33} (D/(5~{\rm kpc})^2$~\ergs\ 
(0.5--8 keV).

%We note that the spectra of some X-ray pulsars, particularly 
%magnetars, are usually modeled as the sum of two spectral components, such 
%as a black body and a power law. Although 
%there is some evidence for a hard excess above 5~keV in the spectrum
%of \wdpulse, the 
%small bandpass provided by our \chandra\ data does not allow us to 
%constrain the parameters of a second continuum component.

\begin{figure}
\centerline{\psfig{file=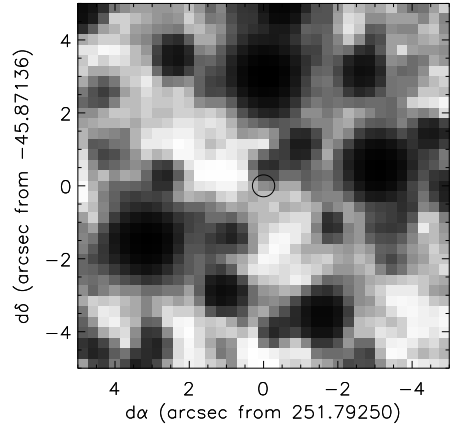,width=0.8\linewidth}}
\caption{
Infrared image a 10\arcsec$\times$10\arcsec\ field around 
\wdpulse. The image was taken with the ESO/NTT in the $K_s$ band, 
and is composed of 10$\times$1.2s images obtained on 2003 June 19 
(P.I.: J. Alves).
The circle denotes the 0\farcs3 uncertainty (90\% confidence)
in the location
of the X-ray source. An object with an intensity of $K_s$=18.4$\pm$0.3 
mag lies 0\farcs5 from the X-ray source. This source is near the 
detection limit, and chance that it is associated with the pulsar
$\approx$1.5\%. The upper limit on the intensity of any object
within the error circle is $K_s$$>$18.5.
}
\label{fig:kimage}
\end{figure}

To evaluate whether \wdpulse\ is an accreting 
system, we searched for an infrared counterpart in $K_s$-band images taken 
with the SofI instrument on the ESO New Technology Telescope 
(Fig.~\ref{fig:kimage}). There was no counterpart to the pulsar within the 
0\farcs3 uncertainty in its location (90\% confidence). The upper limit to its
intensity was $K_s$$\ge$18.5, ruling out a companion with $M$$>$1 
\msun\ \citep{gir02}. For comparison, the faintest stars that have 
contracted onto the main sequence in Westerlund 1 have $M_i$=2\msun\
(Brandner \etal, in prep.). 

%%%%%%%%%%%%%%%%%%%%%%%%%%%%%%%%%%%%%%%%%%%%%%%%%%%%%%%%%%%%%%%
%% K band image

\section{Discussion}

The spectrum and luminosity of \wdpulse\ (Fig.~\ref{fig:spectrum}) 
demonstrate that it is not a conventional radio pulsar. First, the power
lost as the pulsar spins down is too small to produce the observed 
X-ray emission. Assuming 
that it is a 10~km, 1.4\msun\ neutron star ($I \sim 10^{45}$ g cm$^{2}$), 
the upper limit to the period derivative 
($\dot{P} < 2\times10^{-10}$ s s$^{-1}$) implies that the spin-down 
energy is $\dot{E} = 4\pi I\dot{P}/P^3 \le 3\times10^{33}$ 
\ergs. This is similar to the observed X-ray luminosity, whereas 
magnetospheric emission
from a radio pulsar with $\dot{E} \le 10^{35}$~\ergs\ would produce
$L_{\rm X} < 10^{-3}\dot{E}$ \citep{ctw04}. Second, the X-ray emission is 
inconsistent with thermal emission from a young, cooling neutron star. 
Although the
characteristic temperature of the emission ($kT$$\approx$0.6 keV) 
is consistent with the high end of the range expected for cooling neutron 
stars \citep[e.g.,][]{yp04}, 
the luminosity is $100$ times lower than would be expected if the surface of 
the neutron star had a uniform temperature.

In contrast, the X-ray emission is consistent with that of
 highly-magnetized 
($B$$>$$3\times10^{14}$ G), slowly rotating pulsars, which are referred to as
magnetars \citep{dt92}. Known magnetars have spin periods between 5 and 12 s,  
$L_{\rm X} = 10^{33} - 10^{36}$ \ergs, and 
spectra that peak at $\approx$2 keV \citep[e.g.,][]{mer04,wt05}. 
%The only aspect of the magnetar hypothesis that remains 
%to be confirmed is to measure a rapid slowing of the neutron star's rotation,
%which provides an estimate of the magnetic field strength. 
%Magnetars typically
%spin down at rates of between $\dot{P}$$\sim$$10^{-10}$ and 
%$\sim$$10^{-12}$ s s$^{-1}$. Our limit above is not strict enough to 
%detect this rate of spin down. However, a second, similarly-sensitive 
%observation taken over an interval $\Delta t$ a few years long will be 
%able to identify a period change at a rate of 
%$\dot{P} = 3\times10^{-12} (\Delta t/{\rm yr})^{-1}$ s s$^{-1}$.

The X-ray emission is also consistent 
with that from faint, wind-accreting neutron stars, such as A 0535+26 
\citep[$P$=104 s;][]{orl04},
X Per \citep[$P$=837 s;][]{ds98,dm01} and 4U 0115+63 
\citep[$P$=3.6 s;][]{cam01}.
%If this pulsar is accreting,
%it would be the first such system with a pre-main-sequence mass 
%donor.
However, forming a close, accreting binary from a system that initially
contained two stars with $M$$>$40\msun\ (see below) and $M$$<$1\msun\ would 
present a significant challenge to models for producing low-mass 
X-ray binaries \citep[e.g.,][]{pz97}. Therefore, we suspect that \wdpulse\ is 
a magnetar.

It is likely that \wdpulse\ is associated with Westerlund 1,
because Galactic X-ray pulsars with $P$=3--30 s are rare 
\citep[only $\approx$20 known;][]{liu00}.
To evaluate the chance that this pulsar is a random
Galactic object unassociated with Westerlund 1, we attempted
to determine the surface density of slow X-ray pulsars on the sky.
Previous searches for X-ray pulsars were biased toward the most luminous
X-ray sources, so we searched $\approx$350 fields in the Galactic plane 
($-5$$<$$b$$<$5\degree) that were observed with 
\chandra\ and \xmm.  Aside from $\approx$15 known 
examples that were the targets of the relevant observations, we found
no new pulsars with $P$=3--30 s (A. Nechita \etal, in prep). Therefore, 
the chance of discovering a slow X-ray pulsar in any pointing is $\la$0.3\%.
(We note that the $\approx$15 known slow X-ray pulsars 
within the more conservative latitude range of $|b|$$<$1\degree\ have a 
surface density of $\sim$0.03 degree$^{-2}$. The 
ACIS-S FOV is $\approx$0.07 degree$^{-2}$, which also implies that 
the chance of finding a slow pulsar in a \chandra\ observation is 
$\sim$0.2\%.) Moreover, \wdpulse\ lies only 1\farcm7 
from the center of Westerlund 1 (2.3$[D/5~{\rm kpc}]$ pc in 
projection). Based on the spatial distribution of X-ray sources described
above, a source at that location has a $\sim$10\% chance of being 
a random association. We conclude that this pulsar is a member of 
Westerlund 1 with $\sim$99.97\% confidence.

The fact that \wdpulse\ is member of Westerlund 1 places
a lower limit on the initial mass of its progenitor. \citet{cla05} 
have established that the cluster is only 4$\pm$1 Myr 
old, so that the minimum mass of a star that could have undergone a 
supernova is $\approx$40 \msun. This is supported by the 
identification of several O7--O8V stars in the cluster, which have zero-age 
main sequence masses of 34--37 \msun\ (J. S. Clark \etal, in prep.). 
Therefore, \wdpulse\ was produced by a star with $M_i$$>$40\msun.

Previously, only three secure lower limits were obtained on the masses 
of progenitors to neutron stars, all of which were for magnetars. 
First, a shell of HI around 1E 1048.1--5937 was interpreted as 
ISM displaced by the wind of a progenitor 
with $M_i$=30--40\msun \citep{gae05}. Second, SGR 1900+14 was suggested 
to be a member of a star cluster $<$$10$ Myr old \citep{vrb00}, placing a lower
limit on the progenitor mass of $M_{i}$$\ga$20\msun. Finally, SGR 1806--20 
was claimed to be a member of a star cluster that is only $\la$4.5~Myr old
\citep[although see][]{cam05,mcg05},
providing a limit of $M_i$$\ga$50\msun \citep[e.g.,][]{fuc99,fig05}.
Our evidence that \wdpulse\ also descended from a star with 
$M_i$$>$40 \msun\ dispels much of the doubt that the previous results 
represented chance associations. %Moreover, there is a good chance
%that \wdpulse\ is also a magnetar.

These results demonstrate that some stars with 
$M_i$$>$40~\msun\ do not collapse into black holes at the ends of 
their lives, but instead form neutron stars. This implies that 
massive stars can lose $\ge$95\% of their mass either before or during
supernovae. Before supernovae, stars could lose mass through 
strong winds \citep[e.g.,][]{heg03} or be stripped of mass by binary 
companions \citep[e.g.,][]{wl99}. During supernovae, rapidly-rotating 
cores could drive mass away through magneto-hydrodynamic 
winds \citep[e.g.,][]{aw05}. 
%That some $>$40\msun\ stars leave 
%magnetars implies that their cores were rapidly-rotating, as this is 
%required to produce $B$$>$$10^{14}$ G fields (Duncan \& Thomspon 1992).
%The large angular momentum could be a product of binary interactions, or
%it could occur because magnetic breaking does not have time to operate during 
%the short supergiant phase of massive stars 
%\citep{heg05,gae05}. 
To determine the importance of these 
effects, it is necessary to identify additional neutron stars in star clusters
and constrain the masses of their progenitors.

\acknowledgements
We thank B. M. Gaensler and B. Hansen for discussions, L. Hadfield 
for calibrating the photometry in the infrared image, and A. Nechita
for carrying out the archival search for X-ray pulsars. CL and FY-Z 
were supported by a grant from the CXC, MPM through a Hubble Fellowship, and
DP through a Chandra Fellowship.

\end{document}